\documentclass[a4paper,11pt]{article}
\pdfoutput=1

\usepackage{jcappub}
\usepackage[utf8]{inputenc}

\newcommand{\dalembert}{\mathop{}\!\square}
\newcommand{\mathi}{\mathrm{i}}
\newcommand{\mathe}{\mathrm{e}}
\newcommand{\christoffel}[4][]{\,{}{#1}\Gamma^{#2}_{#3 #4}\,}
\newcommand{\eqend}[1]{\,\text{#1}}
\newcommand{\total}{\mathop{}\!\mathrm{d}}
\newcommand{\bigo}[1]{\mathcal{O}\!\left({#1}\right)}
\newcommand{\bra}[1]{{\left\langle{#1}\right\rvert}}
\newcommand{\ket}[1]{{\left\lvert{#1}\right\rangle}}
\newcommand{\expect}[1]{{\left\langle{#1}\right\rangle}}
\newcommand{\bessel}[3]{\mathrm{#1}_{#2}\!\left(#3\right)}

\newcommand{\artanh}{\operatorname{artanh}}
\newcommand{\sgn}{\operatorname{sgn}}
\newcommand{\abs}[1]{{\left\lvert{#1}\right\rvert}}

\DeclareMathAlphabet{\mathbold}{OML}{cmm}{b}{it}
\renewcommand{\vec}[1]{\mathbold{#1}}

\title{Fully renormalized stress tensor correlator in flat space}

\author[a,b]{Markus B. Fröb}
\affiliation[a]{Departament de Física Fonamental, Institut de Ciències del Cosmos (ICC), Universitat de Barcelona (UB), C/ Martí i Franquès 1, 08028 Barcelona, Spain}
\affiliation[b]{Department of Mathematics, University of York, Heslington, York, YO10 5DD, UK}

\emailAdd{mfroeb@ffn.ub.edu}

\abstract{We present a general procedure to renormalize the stress tensor two-point correlation function on a Minkowski background in position space. The method is shown in detail for the case of a free massive scalar field in the standard Minkowski vacuum state, and explicit expressions are given for the counterterms and finite parts, which are in full accordance with earlier results for the massless case. For the general case in position space, only regularized --- but not renormalized --- results have been obtained previously. After a Fourier transformation to momentum space, we also check agreement with a previous calculation there. We generalize our results to general Hadamard states. Furthermore, the proposed procedure can presumably be generalized to the important case of an inflationary spacetime background, where the transition to momentum space is in general not possible.}


\begin{document}
\maketitle
\flushbottom

\section{Introduction}

The fundamental vertex of the perturbative interaction between any type of matter and gravity is given by $h_{ab} T^{ab}$, where $h_{ab}$ is the graviton field (i.e., the perturbation of a fixed background) and $T_{ab}$ is the stress tensor of the matter fields under study. To calculate interaction processes between matter and $n$ gravitons it is therefore necessary to obtain $n$-point functions of the stress tensor, and if gravity interacts with a large number of matter fields those processes will dominate over graviton self-interactions (this fact can be formalised by treating $N$ matter fields in an $1/N$ expansion~\cite{tomboulis1977,hartlehorowitz1981}).

However, because of their extreme smallness quantum effects of gravity are very difficult to detect. Except maybe for analogue models of gravity~\cite{analoguegravity}, the only field where we can realistically hope to see any quantum effects is cosmology, where precision measurements became recently available~\cite{planck2013a,planck2013b,planck2013c}. Quantum fluctuations in the early universe are believed to have seeded structure formation, and a convenient cosmological observable is the power spectrum of primordial gravitational perturbations which manifests itself through temperature fluctuations of the cosmological microwave background. Therefore, obtaining quantum corrections to this power spectrum is an important task, and a basic building block is the two-point function of the stress tensor of matter fields, most importantly of inflaton fields. The general case is of course hard, and so in this paper we restrict ourselves to Minkowski spacetime and hope to generalize this to curved backgrounds in a later work. On the other hand, the proper definition of stress tensor $n$-point functions is also interesting from a mathematical point of view, and having a very concrete procedure may help with a following precise mathematical investigation.

The first obstacle one faces is that the $n$-point functions obtained in the naive way by taking the expectation value of $n$ stress tensor operators are not well defined distributions, since they involve products of fields at the same space-time point. One can improve the situation by defining a renormalized stress tensor operator which has a finite expectation value, or, alternatively, by working with connected $n$-point functions ($n>1$), which gives a finite result as long as all points in the correlation function are not light-like related to each other. However, this is still not a well-defined distribution since there remain divergences for light-like separations which have to be subtracted by counterterms (p.ex., for the two-point function those are terms quadratic in the curvature tensors), and the extraction of those divergences and the exhibition of the finite part is in general a hard problem. Only for the one-point function, i.e., the expectation value of the stress tensor, the renormalization for a generic Hadamard state in a generic curved background is known in full generality~\cite{wald1977,wald1978,horowitz1980,moretti2003,decaninifolacci2005} (this can be used to define the above mentioned renormalized stress tensor operator). For the two-point function, only specific examples are known where one can properly separate divergent and finite parts and perform renormalization, which include scalar fields on a Minkowski background in momentum space~\cite{martinverdaguer2000}, massless conformally coupled scalars on conformally flat backgrounds~\cite{camposverdaguer1994,camposverdaguer1996} and massless minimally coupled scalars in de Sitter space~\cite{parkwoodard2011a}, while higher $n$-point functions become tremendously complicated (see~\cite{osbornpetkou1994,erdmengerosborn1997,arutyunovfrolov1999,cppr2011,stanev2012} for examples in conformal theories).

In this paper, we present a procedure to renormalize the two-point correlation function of the stress tensor in Minkowski spacetime in a relatively easy way. We explain all steps on the concrete example of a massive, minimally coupled scalar field. The method works in position space, and so we expect it to be easier generalizable to curved spaces than momentum space techniques which are ubiquitous in Minkowski space QFT calculations. In the first section, we give a very short summary of Mellin-Barnes integrals which are central to our method, and derive some integral formulas which are needed later.\footnote{Mellin transform techniques have also been used for the calculation of Feynman diagrams in momentum space (see~\cite{smirnovfeynman} for an overview), and in the AdS/CFT correspondence (see p.ex.~\cite{penedones2011,paulos2011,fkprvr2011}).} The next section gives the first step of our procedure, which is the derivation of the (connected) stress tensor two-point function for a massive scalar field. This correlation function is regulated in $n$ dimensions and given in Mellin-Barnes form. In the second step, done in the section afterwards, differential operators are extracted to lower the naive degree of divergence. The resulting expression is what in momentum space would be given as the regularized form before renormalization. In continuation we split the result into divergent and finite parts (as $n \to 4$), and renormalize it by adding local counterterms. At the end of the section, we present the fully renormalized result. The following section generalizes our results to arbitrary Hadamard states (including a concrete simple example), and in the penultimate section the accordance of our results with previously obtained special cases is shown. Subsequently, a possible generalization of the procedure to curved spacetimes is discussed.

We employ the ``+++'' sign convention of \cite{mtw}, and use Latin indices only. The covariant derivative with respect to the general metric $g_{ab}$ is denoted by $\nabla_a$. In flat space, we also use the symbol $\dalembert = \eta^{ab} \partial_a \partial_b$. We use dimensional regularization in $n$ dimensions.

\section{Integral games}

\begin{figure}[h]
\centering
\includegraphics[scale=1.2]{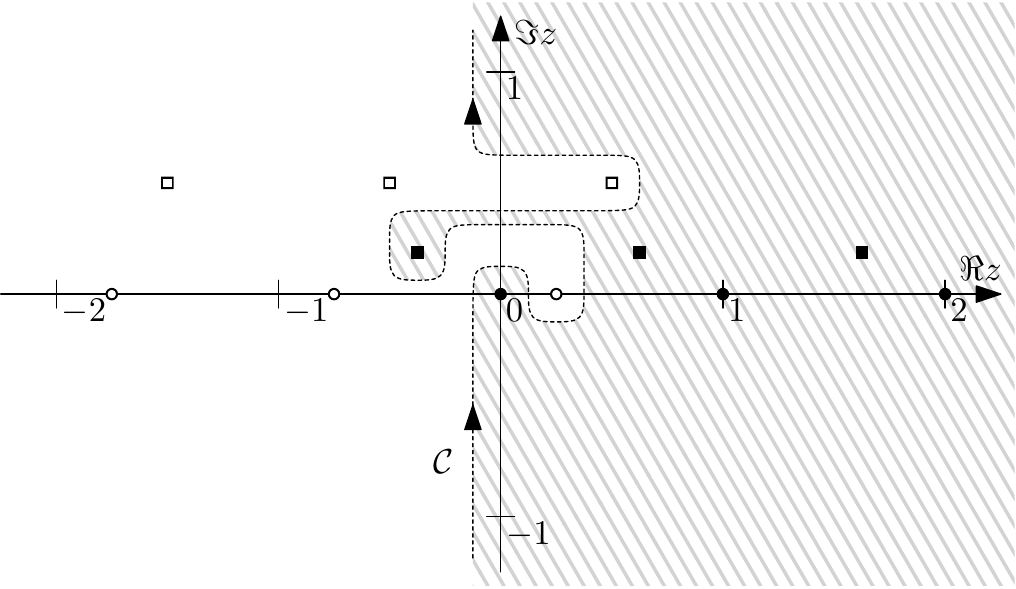}
\caption{The integration path $\mathcal{C}$ for the example Mellin-Barnes integral. It separates the complex plane in a left (unshaded) and a right (shaded) half such that the left poles (shown in white) all lie in the left half and the right poles (shown in black) all lie in the right half.}
\label{mellin_general}
\end{figure}

In this section we explain and derive formulas that will be put to good use in the rest of the paper. All of our results are derived by the judicious application of Mellin-Barnes integrals --- contour integrals in the complex plane which involve products of $\Gamma$ functions. An example is given by
\begin{equation}
\int_{\mathcal{C}} \Gamma(a+z) \Gamma(b+z) \Gamma(c-z) \Gamma(d-z) x^z \frac{\total z}{2\pi\mathi} \eqend{,}
\end{equation}
and the corresponding integration path $\mathcal{C}$ is shown in figure~\ref{mellin_general} for $a=\frac{1}{4}$, $b=\frac{1}{2}(1+\mathi)$, $c=\frac{3}{16}(-2+\mathi)$ and $d=0$.
In general, the integration path goes from $-\mathi\infty$ to $\mathi\infty$, separating left poles (poles of the $\Gamma$ functions of the form $\Gamma(a+z)$) from right poles (poles of the $\Gamma$ functions which are of the form $\Gamma(a-z)$). This is always possible as long as no difference between the $a$'s is an integer, so that no left pole coincides with any right pole. Since the $\Gamma$ functions decay exponentially in imaginary directions, these integrals are well defined, and when they decay also in a real direction $z \to \pm\infty$ they can be evaluated by contour integration, summing the resulting series of residues. An important case is given by Barnes' Lemma~\cite{barnes1908,jantzen2013}
\begin{equation}
\label{barnes_lemma}
\int_{\mathcal{C}} \Gamma(a+z) \Gamma(b+z) \Gamma(c-z) \Gamma(d-z) \frac{\total z}{2\pi\mathi} = \frac{\Gamma(a+c) \Gamma(a+d) \Gamma(b+c) \Gamma(b+d)}{\Gamma(a+b+c+d)} \eqend{.}
\end{equation}
Another case which we need in the sequel is given by
\begin{equation}
\label{mellin_decomposition}
\int_{\mathcal{C}} \frac{\Gamma(\alpha+z) \Gamma(-z)}{\Gamma(\alpha)} x^z \frac{\total z}{2\pi\mathi} = \frac{1}{(1+x)^\alpha} \eqend{,}
\end{equation}
where the contour can be closed to the left if $\abs{x} > 1$ and to the right if $\abs{x} < 1$.

After this very short introduction to Mellin-Barnes integrals, we use them to calculate Fourier transforms that will be important later on. Namely, we need the Fourier transform of the Feynman propagator $1/(p^2+m^2-\mathi \epsilon)$ in the standard Minkowski vacuum state (with the limit $\epsilon \to 0^+$ understood in the sense of distributions, for which we simply write $1/(p^2+m^2-\mathi 0)$). We will start with a general power of the massless case, which is~(\cite{smirnovfeynman}~eq.~A.40, \cite{brychkovprudnikov}~eq.~8.715 converted to our conventions)
\begin{equation}
\label{feynman_massless}
\int \frac{\mathe^{\mathi p x}}{(p^2 - \mathi 0)^\alpha} \frac{\total^n p}{(2\pi)^n} = \mathi \frac{\Gamma\left( \frac{n}{2}-\alpha \right)}{4^\alpha \pi^\frac{n}{2} \Gamma(\alpha)} \frac{1}{(x^2 + \mathi 0)^{\frac{n}{2}-\alpha}} \eqend{.}
\end{equation}
This is a priori only defined for $0 < \Re \alpha < \frac{n}{2}$, but using the reduction formula
\begin{equation}
\label{reduction_power}
(x^2)^{-p} = \frac{1}{2(1-p)(n-2p)} \dalembert (x^2)^{1-p}
\end{equation}
and the fact that the Fourier transform of the d'Alembertian operator is $-p^2$ this can be extended to all complex $\alpha$, except $\alpha = 0$ and $\alpha = \frac{n}{2}$ where the $\Gamma$ functions diverge and the Fourier transform is between a constant and a $\delta$ distribution,
\begin{equation}
\int \mathe^{\mathi p x} \frac{\total^n p}{(2\pi)^n} = \delta^n(x) \eqend{.}
\end{equation}
The massive case then follows by application of~\eqref{mellin_decomposition}~\cite{boosdavydychev1991} and yields
\begin{equation}
\label{feynman_massive}
\begin{split}
\int \frac{\mathe^{\mathi p x}}{(p^2+m^2-\mathi 0)^\alpha} \frac{\total^n p}{(2\pi)^n} &= \mathi \int_\mathcal{C} \frac{(m^2)^z}{(x^2 + \mathi 0)^{\frac{n}{2}-\alpha-z}} \frac{\Gamma\left( \frac{n}{2}-\alpha-z \right) \Gamma(-z)}{4^{\alpha+z} \pi^\frac{n}{2} \Gamma(\alpha)} \frac{\total z}{2\pi\mathi} \\
&= \mathi \frac{m^{n-2\alpha}}{2^{\frac{n}{2}-1+\alpha} \pi^\frac{n}{2} \Gamma(\alpha)} \left( m \sqrt{x^2 + \mathi 0} \right)^{-\frac{n}{2}+\alpha} \bessel{K}{\frac{n}{2}-\alpha}{m \sqrt{x^2 + \mathi 0}} \eqend{,}
\end{split}
\end{equation}
which for $\alpha = 1$ is just the Fourier transform of the Feynman propagator~\cite{bogoliubovshirkov}. Note that in this case the contour may only be closed to the right because in the other half plane the integrand is exponentially diverging as $z \to -\infty$. However, for the following we will only need the Mellin-Barnes representation, and use the notation
\begin{equation}
\label{g_mellinbarnes}
G(x^2) = \int_\mathcal{C} (m^2)^z (x^2)^{z+1-\frac{n}{2}} \frac{\Gamma\left( \frac{n}{2}-1-z \right) \Gamma(-z)}{4^{1+z} \pi^\frac{n}{2}} \frac{\total z}{2\pi\mathi} \eqend{.}
\end{equation}
In a similar manner, one obtains the coordinate space Wightman function using
\begin{equation}
\Theta(p^0) \delta(p^2+m^2) = \frac{\Theta(p^0) \Theta(-p^2)}{2 \pi \mathi} \left( \frac{1}{p^2+m^2-\mathi 0} - \frac{1}{p^2+m^2+\mathi 0} \right)
\end{equation}
and (\cite{brychkovprudnikov}~eq.~8.730 converted to our conventions))
\begin{equation}
\label{wightman_massless}
\int \frac{1}{(x^2 + \mathi 0 \, \sgn t)^\alpha} \mathe^{- \mathi p x} \total^n x = \frac{2^{-2\alpha+n+1} \pi^{\frac{n}{2}+1}}{\Gamma\left( 1 - \frac{n}{2} + \alpha \right) \Gamma(\alpha)} \Theta(p^0) \Theta(-p^2) \frac{1}{(-p^2)^{\frac{n}{2}-\alpha}} \eqend{.}
\end{equation}
It is just given by $G(x^2 + \mathi 0 \, \sgn t)$, so that in coordinate space the various two-point functions only differ by the type of continuation needed to get around the singularity at $x^2 = 0$.

For derivatives, it follows
\begin{equation}
G^{(k)}(x^2) = \int_\mathcal{C} (m^2)^z (x^2)^{z+1-k-\frac{n}{2}} (-1)^k \frac{\Gamma\left( \frac{n}{2}-1-z+k \right) \Gamma(-z)}{4^{1+z} \pi^\frac{n}{2}} \frac{\total z}{2\pi\mathi} \eqend{.}
\end{equation}
For the product of two propagators (and their derivatives), we calculate by shifting the inner integration variable
\begin{equation}
\label{propagator_product}
G^{(k)}(x^2) G^{(l)}(x^2) = \int_\mathcal{C} (m^2)^z (x^2)^{z+2-k-l-n} \frac{(-1)^{k+l}}{4^{2+z} \pi^n} K(k,l,z) \frac{\total z}{2\pi\mathi}
\end{equation}
with
\begin{equation}
\label{kernel_k_def}
\begin{split}
K(k,l,z) &= \int_\mathcal{C} \Gamma\left( \frac{n}{2}-1+y-z+k \right) \Gamma(y-z) \Gamma\left( \frac{n}{2}-1-y+l \right) \Gamma(-y) \frac{\total y}{2\pi\mathi} \\
&= \frac{\Gamma(n-2+k+l-z) \Gamma\left( \frac{n}{2}-1+k-z \right) \Gamma\left( \frac{n}{2}-1+l-z \right) \Gamma(-z)}{\Gamma(n-2+k+l-2z)} \eqend{,}
\end{split}
\end{equation}
where we used Barnes' Lemma~\eqref{barnes_lemma} for the evaluation of the Mellin-Barnes integral. It is important that in the whole process the contour could be chosen such that no pole was traversed. Again, one could evaluate the final integral~\eqref{propagator_product} by closing the contour to the right (since the integrand diverges exponentially as $z \to -\infty$), but it is more useful to keep the Mellin-Barnes representation.

\section{Regularization}

The action of a free massive and minimally coupled scalar field is well known
\begin{equation}
\label{action}
S[g,\phi] = - \frac{1}{2} \int \left( \nabla^a \phi \nabla_a \phi + m^2 \phi^2 \right) \sqrt{-g} \total^n x \eqend{.}
\end{equation}
By taking a functional derivative with respect to $\phi$, we obtain its equation of motion
\begin{equation}
\label{eom_phi}
\left( \nabla^a \nabla_a - m^2 \right) \phi = 0 \eqend{,}
\end{equation}
while by deriving with respect to the metric we obtain the stress tensor
\begin{equation}
\label{stresstensordef}
T_{ab} = \nabla_a \phi \nabla_b \phi - \frac{1}{2} g_{ab} \left( \nabla^c \phi \nabla_c \phi + m^2 \phi^2 \right) \eqend{,}
\end{equation}
which, using the equation of motion for $\phi$~\eqref{eom_phi}, is covariantly conserved $\nabla^a T_{ab} = 0$. This can be written as
\begin{equation}
T_{ab}(x) = \lim_{y \to x} P_{ab}(x,y) \phi(x) \phi(y)
\end{equation}
with
\begin{equation}
\label{stresstensorop}
P_{ab}(x,y) = \left( \delta^c_{(a} \delta^d_{b)} - \frac{1}{2} g_{ab} g^{cd} \right) \nabla^x_c \nabla^y_d - \frac{1}{2} g_{ab} m^2
\end{equation}
and the subscript indicates on which field the derivative acts.

We now specialize to flat space and consider the connected two-point correlation function of the stress tensor
\begin{equation}
\label{connected_def}
\begin{split}
\expect{T_{ab}(x) T_{cd}(x')} &= \bra{0} T_{ab}(x) T_{cd}(x') \ket{0} - \expect{T_{ab}(x)} \expect{T_{cd}(x')} \\
&= \bra{0} \big( T_{ab}(x) - \expect{T_{ab}(x)} \big) \left( T_{cd}(x') - \expect{T_{cd}(x')} \right) \ket{0} \\
\end{split}
\end{equation}
in some state $\ket{0}$ for the scalar field, which we will take to be the standard Minkowski vacuum. The advantage of considering the connected correlation functions is that they are finite as long as the points $x$ and $x'$ are not null (light-like) separated, since the divergences that arise from $\phi^2(x)$ (and its derivatives) in the definition of the stress tensor are absent in the difference $T_{ab}(x) - \expect{T_{ab}(x)}$. However, in the null separation limit $(x-x')^2 \to 0$ the correlation functions still diverge after smearing with test functions; they are not well-defined distributions in four dimensions.

To obtain explicit expressions, we insert the definition of the stress tensor~\eqref{stresstensordef} into the connected two-point function~\eqref{connected_def} and evaluate the expectation values of $k$ matter fields $\phi$ by Wick's formula. 
To simplify the resulting expression, we take advantage of the fact that the operator $P_{mn}(x,y)$~\eqref{stresstensorop} is symmetric under the interchange of $x$ and $y$, and it results
\begin{equation}
\label{npf_in_g}
\expect{T_{ab}(x) T_{cd}(x')} = 2 \lim_{\substack{y\to x\\y'\to x'}} P_{ab}(x,y) P_{cd}(x',y') G(x,x') G(y,y') \eqend{,}
\end{equation}
where
\begin{equation}
G(x,y) = \bra{0} \phi(x) \phi(y) \ket{0} \eqend{.}
\end{equation}
This two-point function only depends on the coordinate difference $x-y$ because of Poincaré invariance of the standard Minkowski vacuum $\ket{0}$, and for simplicity we may set $y=0$. We then get for the two-point function
\begin{equation}
\label{2pf_regularized}
\begin{split}
\expect{T_{ab}(x) T_{cd}(0)} &= \frac{1}{2} \eta_{ab} \eta_{cd} \left[ 16 (x^2 G'')^2 + 16 x^2 G' G'' + 4 ( n-4 + 2 m^2 x^2 ) (G')^2 + m^4 G^2 \right] \\
&\quad+ 8 \eta_{a(c} \eta_{d)b} (G')^2 + 32 x_{(a} \eta_{b)(c)} x_{d)} G' G'' + 32 x_a x_b x_c x_d (G'')^2 \\
&\quad- 4 \left( \eta_{ab} x_c x_d + x_a x_b \eta_{cd} \right) \left[ 4 x^2 (G'')^2 + 4 G' G'' + m^2 (G')^2 \right] \eqend{,}
\end{split}
\end{equation}
where $G'' = \partial^2 G(x,0)/\partial (x^2)^2 = G''(x^2)$ and analogue for $G'$ and $G$. In the above condensed notation, the equation of motion for the scalar field~\eqref{eom_phi} implies
\begin{equation}
4 x^2 G'' + 2 n G' = m^2 G \eqend{,}
\end{equation}
and using it the two-point functions may be simplified somewhat; also covariant conservation can be checked easily.

The attentive reader may have noticed that so far the entire discussion has been in terms of Wightman functions, while for \emph{in-out} calculations Feynman propagators (time-ordered functions) are needed.\footnote{In the \emph{in-in} formalism one even needs the full arsenal, including in addition Dyson (anti-time-ordered) functions and positive and negative frequency Wightman functions.} While they are quite different in momentum space, in coordinate space these various functions differ only by the type of analytic continuation needed to get around the singularity in the propagator as $x^2 \to 0$, as shown in the previous section. When acting with derivatives on the Feynman (or Dyson) propagator, naively additional local terms $\sim \delta^n(x)$ are obtained, so in this case one should find additional local terms in the above expressions for the stress tensor correlation function. However, since a priori the time-ordered stress tensor correlation function is not well defined, there is no reason to keep those local terms. We will instead assume the working hypothesis that the time-ordered two-point function of the stress tensor is obtained by using the Feynman prescription after performing the manipulations in this section and the first half of the next section. This is in accordance with momentum space calculations, where a derivative $\partial_a$ is just multiplication with $\mathi p_a$, and the manipulations in the next section just correspond to a reordering of those $p_a$, without additional terms.

We now put the Mellin-Barnes representation~\eqref{propagator_product} for the product of two propagators (and their derivatives) into the stress tensor two-point function~\eqref{2pf_regularized}. After shifting some integration variables, we obtain
\begin{equation}
\label{2pf_mellin}
\expect{T_{ab}(x) T_{cd}(0)} = \int_\mathcal{C} (m^2)^z (x^2)^{z-n} \frac{1}{4^{z+2} \pi^n} \frac{\Gamma(n-z) \Gamma^2\left( \frac{n}{2}-z \right) \Gamma(-z)}{\Gamma(n+2-2z)} T_{abcd}(z) \frac{\total z}{2\pi\mathi}
\end{equation}
with
\begin{equation}
\label{2pf_mellin_kernel}
\begin{split}
T_{abcd}(z) &= 2 \left[ \eta_{ab} \eta_{cd} (n^2-n-4 - 2z (2n-1) + 4 z^2) + 4 \eta_{a(c} \eta_{d)b} \right] (n+1-2z) (n-2z) \\
&\quad- 16 \frac{x_{(a} \eta_{b)(c)} x_{d)}}{x^2} (n-z) (n-2z) (n+1-2z) \\
&\quad- 4 \frac{\eta_{ab} x_c x_d + x_a x_b \eta_{cd}}{x^2} (n-z) (n-2z) (n-2-2z) (n+1-2z) \\
&\quad+ 8 \frac{x_a x_b x_c x_d}{(x^2)^2} (n+1-z)(n-z) (n-2z)^2 \eqend{.}
\end{split}
\end{equation}
Although this result looks complicated, it is quite simple and has a very nice property. Namely, on physical grounds one expects any stress tensor correlation function to be renormalizable if it fulfils the properties expected from such a function, which is symmetry in the index pairs $(ab)$ and $(cd)$ and conservation. No matter what the exact dependence on the spacetime coordinates looks like, as long as those properties are fulfilled it is conceivable that there is some theory and some quantum state that has this function as the correlation function of its stress tensor. It is now easily checked that conservation is independent of $z$, and so it should be possible to rearrange and renormalize the integrand which is simpler than the whole (integrated) result. That this in fact can be quite simply done is shown in the next section.

\section{Renormalization}

The expression for the two-point function~\eqref{2pf_mellin} derived in the previous section is not yet suited for renormalization. The residue of the first pole in the Mellin-Barnes representation~\eqref{2pf_mellin} is proportional to $(x^2)^{-n}$, which is more singular than the leading singularity in the product of two propagators $[G(x^2)]^2 \sim (x^2)^{2-n}$~\eqref{propagator_product}. This is of course due to the fact that we took derivatives, which in momentum space just correspond to multiplications by $p_a$. To reduce the strength of the singularity, we therefore have to extract differential operators (which in momentum space corresponds to a reordering of the $p_a$), and to reduce the singularity to $(x^2)^{2-n}$ we see that we need fourth-order differential operators. We are led to the exact form of the differential operators we need to extract by the conservation of the stress tensor. Of the five terms appearing in the kernel~\eqref{2pf_mellin_kernel}, conservation entails three relations between them so that we need two independent functions. Furthermore, conservation should be automatically guaranteed by the form of the differential operators. We therefore consider the operators
\begin{equation}
\label{sab_def}
S_{ab} = \partial_a \partial_b - \eta_{ab} \dalembert \eqend{,}
\end{equation}
which fulfil $\partial^a S_{ab} = 0$ identically (i.e., acting on any tensor $T_{m \cdots n}$). An ansatz for the stress tensor two-point function which respects the symmetries is then given by
\begin{equation}
\label{2pf_ansatz_fg}
\expect{T_{ab}(x) T_{cd}(0)} = 2 ( S_{a(c} S_{d)b} - S_{ab} S_{cd} ) f(x^2) + S_{ab} S_{cd} \, g(x^2) \eqend{,}
\end{equation}
where now the functions $f$ and $g$ are unconstrained.\footnote{As a side note for later generalization to curved space backgrounds, we note that $R = S_{ab} h^{ab}$ up to linear order (see appendix~\ref{appendix_expansion}).} To determine $f(x^2)$ and $g(x^2)$, we make for them the ansatz
\begin{equation}
\label{func_f_ansatz}
f(x^2) = \int_\mathcal{C} (m^2)^z (x^2)^{z+2-n} \frac{1}{4^{z+2} \pi^n} \frac{\Gamma(n-z) \Gamma^2\left( \frac{n}{2}-z \right) \Gamma(-z)}{\Gamma(n+2-2z)} F(z) \frac{\total z}{2\pi\mathi}
\end{equation}
and analogue for $g(x^2)$. Plugging this into the general ansatz~\eqref{2pf_ansatz_fg} and comparing with the regularized result~\eqref{2pf_mellin}, we obtain five relations for $F(z)$ and $G(z)$ which are solved by
%
\begin{equation}
\begin{split}
F(z) &= \frac{2z-n}{2 (z+2-n) (z+1-n) (2z+2-n)} \\
G(z) &= \frac{(n-2z)^2}{2 (z+2-n) (z+1-n)} \eqend{,}
\end{split}
\end{equation}
so that we obtain
\begin{equation}
\label{2pf_mellin_diff}
\begin{split}
\expect{T_{ab}(x) T_{cd}(0)} &= ( S_{a(c} S_{d)b} - S_{ab} S_{cd} ) \times \\
&\qquad\times \int_\mathcal{C} (m^2)^z (x^2)^{z+2-n} \frac{\Gamma(n-2-z) \Gamma\left( \frac{n}{2}-1-z \right) \Gamma\left( \frac{n}{2}+1-z \right) \Gamma(-z)}{2^{2z+4} \pi^n \Gamma(n+2-2z)} \frac{\total z}{2\pi\mathi} \\
&\quad+ S_{ab} S_{cd} \int_\mathcal{C} (m^2)^z (x^2)^{z+2-n} \frac{\Gamma(n-2-z) \Gamma^2\left( \frac{n}{2}+1-z \right) \Gamma(-z)}{2^{2z+3} \pi^n \Gamma(n+2-2z)} \frac{\total z}{2\pi\mathi} \eqend{.}
\end{split}
\end{equation}
The contour runs left of $\Re z = 0$, where the first pole is encountered.
\begin{figure}[h]
\centering
\includegraphics[scale=1.2]{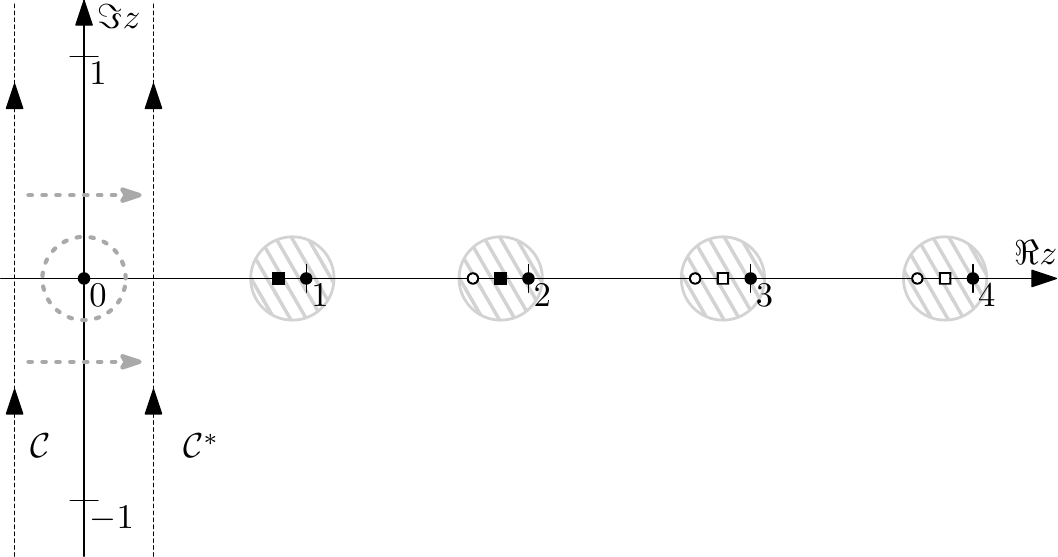}
\caption{The first Mellin-Barnes integral appearing in the stress tensor correlation function. There are simple poles at $z=k$ (black dots), at $z=n-2+k$ (white dots) and at $z=\frac{n}{2}-1+k$ (black squares) and double poles at $z=\frac{n}{2}+1+k$ (white squares), shown here for $n=3.75$. The original contour $\mathcal{C}$ runs left of all poles. We lift it over the pole at $z=0$ to obtain the contour $\mathcal{C}^*$ and the residue of this pole (shown in grey). In the remaining integral over $\mathcal{C}^*$, we can take the limit $n \to 4$, and the poles flow together (shaded circles).}
\label{mellin_contour}
\end{figure}
In four dimensions, the residue of this pole is proportional to $(x^2)^{-2}$ which is still not a well-defined distribution. However, the residues of all other poles are (with the proper Feynman or Wightman prescription) well-defined distributions in four dimensions. We therefore lift the contour over this pole (see figure~\ref{mellin_contour}) to extract the problematic term and take the limit $n\to 4$ in the remaining integral. This leaves us with
\begin{equation}
\label{2pf_mellin_diff_separated}
\begin{split}
\expect{T_{ab}(x) T_{cd}(0)} &= \frac{\Gamma(n-2) \Gamma\left( \frac{n}{2} \right) \Gamma\left( \frac{n}{2}+1 \right)}{16 (n-2) \pi^n \Gamma(n+2)} \left( 2 S_{a(c} S_{d)b} + ( n^2-2n-2 ) S_{ab} S_{cd} \right) (x^2)^{2-n} \\
&\quad+ ( S_{a(c} S_{d)b} - S_{ab} S_{cd} )  \int_{\mathcal{C}^*} (m^2)^z (x^2)^{z-2} \frac{\Gamma(-z) \Gamma(1-z) \Gamma(2-z) \Gamma(3-z)}{2^{2z+4} \pi^4 \Gamma(6-2z)} \frac{\total z}{2\pi\mathi} \\
&\quad+ S_{ab} S_{cd} \int_{\mathcal{C}^*} (m^2)^z (x^2)^{z-2} \frac{\Gamma(-z) \Gamma(2-z) \Gamma^2(3-z)}{2^{2z+3} \pi^4 \Gamma(6-2z)} \frac{\total z}{2\pi\mathi} \eqend{.}
\end{split}
\end{equation}

To treat the problematic term, we apply the reduction formula~\eqref{reduction_power} to obtain
\begin{equation}
\label{x2_m2_reduction}
(x^2)^{2-n} = \frac{1}{2 (n-3) (n-4)} \dalembert (x^2)^{3-n} \eqend{,}
\end{equation}
and the limit $n \to 4$ could be taken to obtain a well-defined distribution, were it not for the explicit power $(n-4)^{-1}$. It is now that the difference between time-ordered functions and Wightman functions becomes important. Namely, for the (massless) Feynman propagator we have~(equation~\eqref{feynman_massless} for $\alpha = 1$)
\begin{equation}
\dalembert \bra{0} \mathcal{T} \phi(x) \phi(0) \ket{0} = \dalembert \left( \frac{\Gamma\left( \frac{n}{2}-1 \right)}{4 \pi^\frac{n}{2}} \frac{1}{(x^2 + \mathi 0)^{\frac{n}{2}-1}} \right) = \mathi \delta^n(x) \eqend{,}
\end{equation}
while for the Wightman function it results
\begin{equation}
\dalembert \bra{0} \phi(x) \phi(0) \ket{0} = \dalembert \left( \frac{\Gamma\left( \frac{n}{2}-1 \right)}{4 \pi^\frac{n}{2}} \frac{1}{(x^2 + \mathi 0 \, \sgn t)^{\frac{n}{2}-1}} \right) = 0 \eqend{.}
\end{equation}
We therefore add to \eqref{x2_m2_reduction} zero in form of $\mu^{n-4} [ (x^2)^{1-\frac{n}{2}} - (x^2)^{1-\frac{n}{2}} ]$ with an arbitrary mass scale $\mu$ and obtain
\begin{equation}
\label{x2_m2_renorm}
\begin{split}
(x^2)^{2-n} &= \frac{1}{2 (n-3) (n-4)} \dalembert \left[ (x^2)^{3-n} - \mu^{n-4} (x^2)^{1-\frac{n}{2}} \right] + \frac{\mu^{n-4}}{2 (n-3) (n-4)} \dalembert (x^2)^{1-\frac{n}{2}} \\
&= - \frac{1}{4} \dalembert \left( \frac{\ln(\mu^2 x^2)}{x^2} \right) + \frac{\mu^{n-4}}{2 (n-3) (n-4)} \dalembert (x^2)^{1-\frac{n}{2}} + \bigo{n-4} \eqend{,}
\end{split}
\end{equation}
which for the Feynman propagator reads
\begin{equation}
\label{feynman_renormalized}
(x^2 + \mathi 0)^{2-n} = - \frac{1}{4} \dalembert \left( \frac{\ln(\mu^2 x^2 + \mathi 0)}{x^2 + \mathi 0} \right) + \mathi \pi^2 \left( \frac{2}{n-4} - 2 + \gamma + \ln \pi + 2 \ln \mu \right) \delta^n(x) + \bigo{n-4} \eqend{,}
\end{equation}
while for the Wightman function we just have
\begin{equation}
(x^2 + \mathi 0 \, \sgn t)^{2-n} = - \frac{1}{4} \dalembert \left( \frac{\ln(\mu^2 x^2 + \mathi 0 \, \sgn t)}{x^2 + \mathi 0 \, \sgn t} \right) + \bigo{n-4} \eqend{.}
\end{equation}
Those formulas just effect the extension of the problematic term to the diagonal $x^2 = 0$. That is, they coincide with the original term for all $x^2 \neq 0$, but are well-defined distributions (in four dimensions) also when smearing with test functions which have support at $x^2 = 0$. The procedure described above is just dimensional regularization and renormalization in coordinate space~\cite{kazakov1983,smirnovfeynman,keller2010}, and the connection with the usual momentum space techniques is described in appendix~\ref{appendix_regren}.

Summarizing, we can decompose the stress tensor two-point function into a singular and regular part, where the singular part is local and only appears for the Feynman (and Dyson) propagator. The singular terms have to be subtracted with counterterms, while the regular part is the renormalized stress tensor correlation function. Schematically, we write for the various two-point functions ($\mathcal{T}^{-1}$ stands for anti-time ordering)
\begin{equation}
\label{tt_result}
\begin{split}
\expect{\mathcal{T} T_{ab}(x) T_{cd}(0)} &= \mathi \expect{T_{ab} T_{cd}}^\text{sing}(x) + \expect{T_{ab} T_{cd}}^\text{reg}(x^2 + \mathi 0) \\
\expect{T_{ab}(x) T_{cd}(0)} &= \expect{T_{ab} T_{cd}}^\text{reg}(x^2 + \mathi 0 \, \sgn t) \\
\expect{T_{cd}(0) T_{ab}(x)} &= \expect{T_{ab} T_{cd}}^\text{reg}(x^2 - \mathi 0 \, \sgn t) \\
\expect{\mathcal{T}^{-1} T_{ab}(x) T_{cd}(0)} &= - \mathi \expect{T_{ab} T_{cd}}^\text{sing}(x) + \expect{T_{ab} T_{cd}}^\text{reg}(x^2 - \mathi 0) \eqend{,}
\end{split}
\end{equation}
where the regular part is given by
\begin{equation}
\label{tt_regular}
\begin{split}
\expect{T_{ab} T_{cd}}^\text{reg}(x^2) &= ( S_{a(c} S_{d)b} - S_{ab} S_{cd} ) \Bigg[ - \dalembert \left( \frac{\ln(\mu^2 x^2)}{3840 \pi^4 x^2} \right) \\
&\qquad+ \int_{\mathcal{C}^*} (m^2)^z (x^2)^{z-2} \frac{\Gamma(-z) \Gamma(1-z) \Gamma(2-z)}{512 \pi^\frac{7}{2} \Gamma\left( \frac{7}{2} - z \right)} \frac{\total z}{2\pi\mathi} \Bigg] \\
&\quad+ S_{ab} S_{cd} \left[ - \dalembert \left( \frac{\ln(\mu^2 x^2)}{960 \pi^4 x^2} \right) + \int_{\mathcal{C}^*} (m^2)^z (x^2)^{z-2} \frac{\Gamma(-z) \Gamma(2-z) \Gamma(3-z)}{256 \pi^\frac{7}{2} \Gamma\left( \frac{7}{2} - z \right)} \frac{\total z}{2\pi\mathi} \right]
\end{split}
\end{equation}
and the singular part reads
\begin{equation}
\begin{split}
\expect{T_{ab} T_{cd}}^\text{sing}(x) &= \frac{1}{960 \pi^2} \left( \frac{2}{n-4} - \frac{46}{15} - \gamma - \ln \pi + 2 \ln \mu \right) ( S_{a(c} S_{d)b} - S_{ab} S_{cd} ) \, \delta^n(x) \\
&\quad+ \frac{1}{240 \pi^2} \left( \frac{2}{n-4} - \frac{47}{30} - \gamma - \ln \pi + 2 \ln \mu \right) S_{ab} S_{cd} \, \delta^n(x) \eqend{.}
\end{split}
\end{equation}
We now show that those singular terms can be subtracted by known counterterms which are quadratic in curvature tensors. For this we need that up to surface terms and to linear order in the perturbation $h_{ab}$ we have
\begin{equation}
\begin{split}
\int h^{ab} S_{ad} S_{bc} h^{cd} \total^4 x &= \int R^{abcd} R_{abcd} \sqrt{-g} \total^4 x = 4 \int R^{ab} R_{ab} \sqrt{-g} \total^4 x - \int R^2 \sqrt{-g} \total^4 x \\
\int h^{ab} S_{ab} S_{cd} h^{cd} \total^4 x &= \int R^2 \sqrt{-g} \total^4 x \eqend{,}
\end{split}
\end{equation}
which can be easily proven using the expansions given in appendix~\ref{appendix_expansion}. We therefore obtain
\begin{equation}
\label{counterterms}
\begin{split}
&480 \pi^2 \int h^{ab}(x) \expect{T_{ab} T_{cd}}^\text{sing}(x-x') h^{cd}(x') \total^4 x \total^4 x' = \\
&\quad- 6 \int \! R^{ab} R_{ab} \sqrt{-g} \total^4 x + \left( \frac{2}{n-4} - \frac{1}{15} - \gamma - \ln \pi + 2 \ln \mu \right) \int \! \left( 2 R^{ab} R_{ab} + R^2 \right) \sqrt{-g} \total^4 x \eqend{.}
\end{split}
\end{equation}
The infinite part of this expression coincides with well-known results~\cite{thooftveltman1974,fordwoodard2005}.

\section{General Hadamard states}

A general state of a free quantum field is called Hadamard if near coincidence its two-point function is of the form
\begin{equation}
G(x,x') \sim - \frac{u(x,x')}{4 \pi^2 (x-x')^2} + v(x,x') \ln ((x-x')^2) + w(x,x') \eqend{,}
\end{equation}
where $u(x,x')$, $v(x,x')$ and $w(x,x')$ are smooth functions with $u(x,x) = 1$~\cite{hadamard,dewittbrehme1960}. Hadamard states have a lot of nice physical properties (for example, the Hadamard form is preserved under Cauchy evolution, and vacuum states in static spacetimes are Hadamard), and it has been shown in examples that non-Hadamard states have unpleasant behaviour (such as an infinite stress tensor expectation value with usual renormalization procedures)~\cite{fullingsweenywald1978,rumpf1981,fullingnarcowichwald1981,najmiottewill1984,weiss1986,kaywald1991,radzikowski1996,hollandswald2002,hollandswald2010,fewsterverch2012}. In particular, the standard Minkowski vacuum state $\ket{0}$ is a Hadamard state, which follows by expanding the two-point function~\eqref{feynman_massive} in four dimensions near $x^2 = 0$,
\begin{equation}
G_0(x^2) = - \frac{m}{(2\pi)^2 \sqrt{x^2}} \bessel{K}{1}{m \sqrt{x^2}} \sim - \frac{1}{4 \pi^2 x^2} + \frac{m^2}{16 \pi^2} \left( 1 - 2 \gamma - \ln (m^2 x^2/4) \right) + \bigo{x^2} \eqend{.}
\end{equation}
Since $v(x)$ is fully determined by the geometry of the underlying spacetime, two Hadamard states differ only by a smooth function. Especially, the two-point function in any Hadamard state $\ket{\text{H}}$ can be written as
\begin{equation}
G_\text{H}(x,x') = G_0((x-x')^2) + w_\text{H}(x,x') \eqend{,}
\end{equation}
where $w_\text{H}(x,x')$ is a smooth function. If we therefore calculate the stress tensor correlation function in a general Hadamard state, we obtain (using~\eqref{npf_in_g} and $(\partial G_0) w_\text{H} = \partial (G_0 w_\text{H}) - G_0 (\partial w_\text{H})$)
\begin{equation}
\begin{split}
\expect{T_{ab}(x) T_{cd}(x')}_H &= \expect{T_{ab}(x) T_{cd}(x')}_0 + \frac{1}{2} m^4 \eta_{ab} \eta_{cd} w_\text{H} w_\text{H} + m^4 \eta_{ab} \eta_{cd} G_0 w_\text{H} \\
&\quad- m^2 \eta_{ab} \eta_{cd}^{pq} \left[ \left( \partial^{x'}_p w_\text{H} \right) \left( \partial^{x'}_q w_\text{H} \right) + 2 \partial^{x'}_q \left( G_0 \partial^{x'}_p w_\text{H} \right) - 2 G_0 \partial^{x'}_q \partial^{x'}_p w_\text{H} \right] \\
&\quad- \eta_{cd} m^2 \eta_{ab}^{kl} \left[ \left( \partial^x_k w_\text{H} \right) \left( \partial^x_l w_\text{H} \right) + 2 \partial^x_l \left( G_0 \partial^x_k w_\text{H} \right) - 2 G_0 \partial^x_l \partial^x_k w_\text{H} \right] \\
&\quad+ 2 \eta_{ab}^{kl} \eta_{cd}^{pq} \left( \partial^x_k \partial^{x'}_p w_\text{H} \right) \left( \partial^x_l \partial^{x'}_q w_\text{H} \right) \\
&\quad+ 4 \eta_{ab}^{kl} \eta_{cd}^{pq} \left[ \partial^x_k \partial^{x'}_p \left( G_0 \partial^x_l \partial^{x'}_q w_\text{H} \right) + G_0 \partial^x_k \partial^x_l \partial^{x'}_p \partial^{x'}_q w_\text{H} \right] \\
&\quad- 4 \eta_{ab}^{kl} \eta_{cd}^{pq} \left[ \partial^x_k \left( G_0 \partial^x_l \partial^{x'}_p \partial^{x'}_q w_\text{H} \right) + \partial^{x'}_p \left( G_0 \partial^x_k \partial^x_l \partial^{x'}_q w_\text{H} \right) \right] \\
\end{split}
\end{equation}
with the short-hand notation
\begin{equation}
\eta_{ab}^{kl} = \delta^k_{(a} \delta^l_{b)} - \frac{1}{2} \eta_{ab} \eta^{kl} \eqend{,}
\end{equation}
and $w_\text{H}$ and $G_0$ depend on $x$ and $x'$. If we replace $\expect{T_{ab}(x) T_{cd}(x')}_0$ by its renormalized value~\eqref{tt_regular}, this is already a well-defined distribution, because derivatives of a smooth function (of $w_\text{H}$) are still smooth and a distribution ($G_0$) can be multiplied by a smooth function to yield another well-defined distribution. Especially, the counterterms~\eqref{counterterms} which one needs to subtract the divergences as $n \to 4$ are the same for all Hadamard states.

As a simple example, with the scalar field decomposed in the standard creation and annihilation operators~\cite{peskinschroeder,bogoliubovshirkov}, we consider the states $\ket{\alpha,\vec{k}} = \exp[ \mathi \alpha ( a^\dagger_\vec{k} + a_\vec{k} ) ] \ket{0}$ for real $\alpha$ and in the massless case $m=0$. Using the Baker-Campbell-Hausdorff formula~\cite{bch}, we see that they are normalized, and an exercise in commutation relations\footnote{Derive first $[ ( a^\dagger_\vec{q} + a_\vec{q} )^k, \phi(x) ] = - \mathi k \left( 2 / \abs{\vec{q}} \right)^\frac{1}{2} [ \sin(q x) ]_{q^0 = \abs{\vec{q}}} ( a^\dagger_\vec{q} + a_\vec{q} )^{k-1}$ and then commute the exponential with the fields.} then gives the two-point function in such a state which is
\begin{equation}
\bra{\alpha,\vec{k}} \phi(x) \phi(x') \ket{\alpha,\vec{k}} = \bra{0} \phi(x) \phi(x') \ket{0} + \frac{2 \alpha^2}{\abs{\vec{k}}} [ \sin(k x) \sin(k x') ]_{k^0 = \abs{\vec{k}}}
\end{equation}
The smooth difference between the Minkowski vacuum two-point function and this special Hadamard state is therefore given by
\begin{equation}
w_{\alpha,\vec{k}}(x,x') = \frac{2 \alpha^2}{\abs{\vec{k}}} \sin(\vec{k} \vec{x} - \abs{\vec{k}} t) \sin(\vec{k} \vec{x}' - \abs{\vec{k}} t') \eqend{,}
\end{equation}
and its renormalized stress tensor two-point function reads
\begin{equation}
\begin{split}
\expect{T_{ab}(x) T_{cd}(x')}^\text{reg}_{\alpha,\vec{k}} &= - \left( S_{a(c} S_{d)b} + 3 S_{ab} S_{cd} \right) \dalembert \left( \frac{\ln(\mu^2 (x-x')^2)}{3840 \pi^4 (x-x')^2} \right) \\
&\quad+ 8 \alpha^4 \frac{k_a k_b k_c k_d}{\vec{k}^2} \cos^2(k x) \cos^2(k x') \\
&\quad+ \frac{2 \alpha^2}{\pi^2 \abs{\vec{k}}} \cos(k x) \cos(k x') \eta_{ab}^{kl} \eta_{cd}^{pq} k_k k_p \, \partial_l \partial_q (x-x')^{-2} \eqend{,}
\end{split}
\end{equation}
where $k^0 = \abs{\vec{k}}$ is understood.

\section{Comparison with earlier results}

In earlier calculations, only partial results were obtained. The general regularized case (including a general coupling to the curvature $\xi R \phi^2$ in the action) was considered in~\cite{chohu2011}, which coincides with our result~\eqref{2pf_regularized} if one replaces $G(x^2)$ with the explicit form~\eqref{feynman_massive}. However, no renormalization was performed.

In~\cite{fordroman2005,fordwoodard2005} renormalization was performed, but only for the massless case. In this limit, the Mellin-Barnes integral in the renormalized result~\eqref{tt_regular} vanishes because the modified contour $\mathcal{C}^*$ has $\Re z > 0$, so that we are left with the first term only
\begin{equation}
\label{tt_regular_massless}
\expect{T_{ab} T_{cd}}^\text{reg}(x^2) \to - \frac{1}{3840 \pi^4} ( S_{a(c} S_{d)b} + 3 S_{ab} S_{cd} ) \dalembert \left( \frac{\ln(\mu^2 x^2)}{x^2} \right) \quad (m\to 0) \eqend{.}
\end{equation}
By inserting the explicit definition of the operators $S_{ab}$~\eqref{sab_def}, one obtains exactly the result in~\cite{fordwoodard2005}, whereas to show agreement with the result of~\cite{fordroman2005} we must extract another d'Alembertian operator
\begin{equation}
\dalembert \ln^2(\mu^2 x^2) = \frac{8 \left[ \ln(\mu^2 x^2) + 1 \right]}{x^2} \eqend{.}
\end{equation}
Since there only the Wightman function was considered where no additional local terms arise, we obtain
\begin{equation}
\dalembert \dalembert \ln^2(\mu^2 x^2 + \mathi 0 \, \sgn t) = 8 \dalembert \left( \frac{\ln(\mu^2 x^2 + \mathi 0 \, \sgn t)}{x^2 + \mathi 0 \, \sgn t} \right)
\end{equation}
and find full agreement with our result.

The only fully renormalized result for general mass and coupling to curvature was derived in~\cite{martinverdaguer2000}, but in momentum space where the calculation is cumbersome as stated by the authors themselves. To compare this with our coordinate space result, we need to take the Fourier transform for all the different prescriptions. Since the integration over the contour $\mathcal{C}^*$ in the finite part~\eqref{tt_regular} is absolutely convergent, we can exchange this integration with the Fourier transformation and Fourier transform the integrand. For the Feynman prescription we may use the formula~\eqref{feynman_massless}, while for the Wightman prescription formula~\eqref{wightman_massless} is the right one. To fully transform the renormalized result~\eqref{tt_regular}, we need also the Fourier transforms of $\ln(x^2)/x^2$ for the different prescriptions. This can be obtained by writing
\begin{equation}
\frac{\ln(x^2)}{x^2} = \lim_{\epsilon \to 0} \frac{1}{\epsilon} \left[ \frac{1}{(x^2)^{1+\epsilon}} - \frac{1}{(x^2)^{1+2\epsilon}} \right] \eqend{,}
\end{equation}
and by using~\eqref{feynman_massless} and \eqref{wightman_massless} we calculate
\begin{equation}
\begin{split}
\int \frac{\ln (x^2 + \mathi 0)}{(x^2 + \mathi 0)} \mathe^{-\mathi p x} \total^4 x &= \frac{4 \mathi \pi^2}{(p^2 - \mathi 0)} \left[ 2 \gamma - \ln 4 + \ln(p^2 - \mathi 0) \right] \\
\int \frac{\ln (x^2 + \mathi 0 \, \sgn t)}{(x^2 + \mathi 0 \, \sgn t)} \mathe^{-\mathi p x} \total^4 x &= 8 \pi^3 \Theta(p^0) \Theta(-p^2) \frac{1}{p^2} \eqend{.}
\end{split}
\end{equation}
Combining all of the above, for the Feynman prescription we obtain from~\eqref{tt_regular}
\begin{equation}
\label{tt_regular_feynman}
\begin{split}
&\int \expect{T_{ab} T_{cd}}^\text{reg}(x^2+\mathi 0) \mathe^{-\mathi p x} \total^4 x = \\
&\quad+ \frac{\mathi}{960 \pi^2} ( P_{a(c} P_{d)b} + 3 P_{ab} P_{cd} ) \left( 2 \gamma - \ln 4 + \ln(p^2 - \mathi 0) - \ln(\mu^2) \right) \\
&\quad- \frac{\mathi}{512 \pi^\frac{3}{2}} ( P_{a(c} P_{d)b} - P_{ab} P_{cd} ) \int_{\mathcal{C}^*} \sigma^z \frac{\Gamma(z) \Gamma(-z) \Gamma(1-z)}{\Gamma\left( \frac{7}{2} - z \right)} \frac{\total z}{2\pi\mathi} \\
&\quad- \frac{\mathi}{256 \pi^\frac{3}{2}} P_{ab} P_{cd} \int_{\mathcal{C}^*} \sigma^z \frac{\Gamma(z) \Gamma(-z) \Gamma(3-z)}{\Gamma\left( \frac{7}{2} - z \right)} \frac{\total z}{2\pi\mathi} \eqend{,}
\end{split}
\end{equation}
where the operators $P_{ab} = p^2 \eta_{ab} - p_a p_b$ are the Fourier transforms of the operators $S_{ab}$~\eqref{sab_def} and we defined
\begin{equation}
\sigma = \frac{4 m^2}{p^2 - \mathi 0} \eqend{.}
\end{equation}
Note that because of the product of $\Gamma$ functions $\Gamma(z) \Gamma(-z)$ in the numerator, the standard Mellin-Barnes contour $\mathcal{C}$ is not defined for those integrals, but the modified contour $\mathcal{C}^*$ (see figure~\ref{mellin_contour}) works. We may then close the contour to the left if $\abs{\sigma} > 1$ or to the right if $\abs{\sigma} < 1$ and sum the residues to obtain
\begin{equation}
\label{tt_regular_feynman_sum}
\begin{split}
&\int \expect{T_{ab} T_{cd}}^\text{reg}(x^2+\mathi 0) \mathe^{-\mathi p x} \total^4 x = \\
&\quad+ \frac{\mathi}{480 \pi^2} ( P_{a(c} P_{d)b} - P_{ab} P_{cd} ) \left[ \frac{46}{30} + \gamma + \ln\left( \frac{m}{2\mu} \right) + \frac{1}{7} \sum_{k=0}^\infty \frac{\Gamma\left( \frac{9}{2} \right) \Gamma(k+1)}{\Gamma\left( k + \frac{9}{2} \right)} (-1)^k \sigma^{-k-1} \right] \\
&\quad+ \frac{\mathi}{240 \pi^2} P_{ab} P_{cd} \left[ \frac{47}{30} + 2 \gamma + 2 \ln\left( \frac{m}{2\mu} \right) + \frac{1}{7} \sum_{k=0}^\infty \frac{\Gamma\left( \frac{9}{2} \right) \Gamma(k+4)}{(k+1) \Gamma\left( k+\frac{9}{2} \right)} (-1)^k \sigma^{-k-1} \right] \eqend{.}
\end{split}
\end{equation}
With hindsight, we rewrite this as a sum over $1+\sigma$ using
\begin{equation}
\sum_{k=0}^\infty a_k \sigma^{-k-1} = \sum_{m=0}^\infty \frac{(1+\sigma)^m}{m!} \left( \sum_{k=0}^\infty a_k \frac{(k+m)!}{k!} (-1)^{k+1} \right) \eqend{.}
\end{equation}
The sums in the expression~\eqref{tt_regular_feynman_sum} then can be done in terms of Gauß' hypergeometric function which reduces to even simpler functions for the cases considered, and it results
\begin{equation}
\begin{split}
\sum_{k=0}^\infty \frac{\Gamma\left( \frac{9}{2} \right) \Gamma(k+1)}{\Gamma\left( k + \frac{9}{2} \right)} (-1)^k \sigma^{-k-1} &= - \sum_{m=0}^\infty \frac{(1+\sigma)^m}{m!} \left( \sum_{k=0}^\infty \frac{\Gamma\left( \frac{9}{2} \right) \Gamma(k+m+1)}{\Gamma\left( k + \frac{9}{2} \right)} \right) \\
&= - \frac{7}{15} [ 3 + 5 (1+\sigma) + 15 (1+\sigma)^2 ] + 7 (1+\sigma)^\frac{5}{2} \artanh \sqrt{1+\sigma}
\end{split}
\end{equation}
and
\begin{equation}
\begin{split}
\sum_{k=0}^\infty \frac{\Gamma\left( \frac{9}{2} \right) \Gamma(k+4)}{(k+1) \Gamma\left( k+\frac{9}{2} \right)} (-1)^k \sigma^{-k-1} &= - \frac{7}{60} ( 94 - 45 \sigma + 45 \sigma^2 ) \\
&\quad+ \frac{7}{4} \sqrt{1+\sigma} (8 - 4 \sigma + 3 \sigma^2) \artanh \sqrt{1+\sigma} \eqend{.}
\end{split}
\end{equation}
For the Fourier transform of the Feynman prescription stress tensor correlation function~\eqref{tt_regular_feynman_sum} we therefore obtain
\begin{equation}
\label{tt_regular_feynman_result}
\begin{split}
&\int \expect{T_{ab} T_{cd}}^\text{reg}(x^2+\mathi 0) \mathe^{-\mathi p x} \total^4 x = \\
&\quad+ \frac{\mathi}{480 \pi^2} ( P_{a(c} P_{d)b} - P_{ab} P_{cd} ) \left[ \gamma + \ln\left( \frac{m}{2\mu} \right) - \frac{1}{3} \sigma ( 7 + 3 \sigma ) + (1+\sigma)^\frac{5}{2} \artanh \sqrt{1+\sigma} \right] \\
&\quad+ \frac{\mathi}{960 \pi^2} P_{ab} P_{cd} \left[ 8 \gamma + 8 \ln\left( \frac{m}{2\mu} \right) + 3 \sigma ( 1-\sigma ) + \sqrt{1+\sigma} (8 - 4 \sigma + 3 \sigma^2) \artanh \sqrt{1+\sigma} \right] \eqend{.}
\end{split}
\end{equation}

The result of~\cite{martinverdaguer2000} is given for the minimally coupled case and for the Feynman prescription as (expressed in our notation)
\begin{equation}
\begin{split}
\expect{T_{ab} T_{cd}}^\text{reg}(x^2 + \mathi 0) &= \frac{\mathi}{4 \pi^2} \Bigg[ \frac{1}{180} \left( 3 S_{a(c} S_{d)b} - S_{ab} S_{cd} \right) \int (1+\sigma)^2 \phi(p^2) \mathe^{\mathi p x} \frac{\total^4 p}{(2\pi)^4} \\
&\qquad\quad+ \frac{1}{72} S_{ab} S_{cd} \int (\sigma-2)^2 \phi(p^2) \mathe^{\mathi p x} \frac{\total^4 p}{(2\pi)^4} \\
&\qquad\quad- m^2 \left( \frac{2}{135} \left( 3 S_{a(c} S_{d)b} - S_{ab} S_{cd} \right) + \frac{1}{27} S_{ab} S_{cd} \right) \int \mathe^{\mathi p x} \frac{1}{p^2} \frac{\total^4 p}{(2\pi)^4} \Bigg]
\end{split}
\end{equation}
up to local terms, which anyway depend on the exact renormalization scheme used. In their appendix we find
\begin{equation}
\phi(p^2) = - 2 + \sqrt{1+\sigma} \ln\left( \frac{\sqrt{1+\sigma} + 1}{\sqrt{1+\sigma} - 1} \right) = - 2 + 2 \sqrt{1+\sigma} \artanh \sqrt{1+\sigma} \eqend{,}
\end{equation}
and so their result for the Feynman prescription is (up to local terms)
\begin{equation}
\begin{split}
&\int \expect{T_{ab} T_{cd}}^\text{reg}(x^2+\mathi 0) \mathe^{-\mathi p x} \total^4 x = \\
&\quad+ \frac{\mathi}{120 \pi^2} ( P_{a(c} P_{d)b} - P_{ab} P_{cd} ) \left[ - \frac{1}{3} \sigma (7+3\sigma) + (1+\sigma)^\frac{5}{2} \artanh \sqrt{1+\sigma} \right] \\
&\quad+ \frac{\mathi}{240 \pi^2} P_{ab} P_{cd} \left[ 3 \sigma (1-\sigma) + (8-4\sigma+3\sigma^2) \sqrt{1+\sigma} \artanh \sqrt{1+\sigma} \right] \eqend{.}
\end{split}
\end{equation}
Up to a global factor of $4$, this coincides exactly with our result~\eqref{tt_regular_feynman_result}.

For the Wightman prescription, we do the same steps to obtain
\begin{equation}
\label{tt_regular_wightman_result}
\begin{split}
\int \expect{T_{ab} T_{cd}}^\text{reg}(x^2 + \mathi 0 \, \sgn t) \mathe^{-\mathi p x} \total^4 x &= \frac{1}{480 \pi} ( P_{a(c} P_{d)b} - P_{ab} P_{cd} ) \Theta(p^0) \Theta(-p^2) (1+\sigma)^\frac{5}{2} \\
&\quad+ \frac{1}{960 \pi} P_{ab} P_{cd} \Theta(p^0) \Theta(-p^2) \sqrt{1+\sigma} (8 - 4 \sigma + 3 \sigma^2) \eqend{.}
\end{split}
\end{equation}
For this prescription, the result of~\cite{martinverdaguer2000} reads (expressed in our notation)
\begin{equation}
\begin{split}
\expect{T_{ab} T_{cd}}^\text{reg}(x^2 + \mathi 0 \, \sgn t) &= \frac{\pi^2}{45} \left( 3 S_{a(c} S_{d)b} - S_{ab} S_{cd} \right) \int (1+\sigma)^2 I(p) \mathe^{\mathi p x} \frac{\total^4 p}{(2\pi)^4} \\
&\quad+ \frac{\pi^2}{18} S_{ab} S_{cd} \int (\sigma-2)^2 I(p) \mathe^{\mathi p x} \frac{\total^4 p}{(2\pi)^4} \eqend{,}
\end{split}
\end{equation}
where
\begin{equation}
I(p) = \frac{1}{8 (2\pi)^3} ( 1 - \sgn p^0 ) \Theta(-p^2-4m^2) \sqrt{1+\sigma} = \frac{1}{32 \pi^3} \Theta(-p^0) \Theta(-p^2) \Theta(1+\sigma) \sqrt{1+\sigma} \eqend{,}
\end{equation}
and therefore they have
\begin{equation}
\begin{split}
&\int \expect{T_{ab} T_{cd}}^\text{reg}(x^2 + \mathi 0 \, \sgn t) \mathe^{-\mathi p x} \total^4 x = \\
&\quad+ \frac{1}{480 \pi} ( P_{a(c} P_{d)b} - P_{ab} P_{cd} ) \Theta(-p^0) \Theta(-p^2) \Theta(1+\sigma) (1+\sigma)^\frac{5}{2} \\
&\quad+ \frac{1}{960 \pi} P_{ab} P_{cd} \Theta(-p^0) \Theta(-p^2) \Theta(1+\sigma) ( 8 - 4 \sigma + 3 \sigma^2 ) \sqrt{1+\sigma} \eqend{,}
\end{split}
\end{equation}
in full accordance with our result~\eqref{tt_regular_wightman_result} except for the sign of $p^0$ (which probably is only a typographic error) and the additional Heaviside function $\Theta(1+\sigma)$.

In summary, we may conclude that up to minor differences our result is in full agreement with earlier findings.

\section{Conclusions}

We have presented an explicit procedure to obtain renormalized two-point functions of stress tensors directly in coordinate space, and have given the full renormalized result~\eqref{tt_result} for the case of a minimally coupled scalar field in Minkowski spacetime. Our result provides an important check on earlier partial results and results obtained in momentum space. This result can now be used to determine quantum matter corrections to the background geometry which give rise to a variety of interesting effects, also in curved spacetime~\cite{ford1994,fordsvaiter1996,fordwu2000,amm2003,phillipshu2003,sinharavalhu2003,borgmanford2004,thompsonford2006,ford2007,fordwu2007,wungford2007,roura2007,huroura2007a,huroura2007b,thompsonford2008,huverdaguer2008,amm2009,fordmiaongwoodardwu2010,fewsterfordroman2012,frv2012,fprv2013}.

Our method rests on two points: first, a linearization formula~\eqref{propagator_product} for the product of two propagators and their derivatives, representing them as a Mellin-Barnes integral; and second, the extraction of differential operators to bring the correlation function into a form well-suited for renormalization~\eqref{2pf_mellin_diff}. The method works directly in coordinate space and so should be more easily generalizable to curved spacetimes where no Fourier transform is defined, which is especially important in the cosmological context. In fact, for the relevant case of a de Sitter background, a similar linearization formula for the product of two (undifferentiated) propagators has been derived~\cite{marolfmorrison2010a,marolfmorrison2010b,marolfmorrison2011} and used in some applications~\cite{higuchimarolfmorrison2010,morrison2013} (see also~\cite{hollands2013,hollands2012}). Of course, the linearization formula derived in this article depends on the fact that the two-point function of a scalar field state invariant under the Poincaré group only depends on the invariant distance $(x-x')^2$. For a generalization to curved spacetimes, this will not the case; however, say for a FLRW universe with scale factor $a(\eta)$ depending on the conformal time $\eta$, is is conceivable that a similar Mellin-Barnes representation exists with the kernel additionally depending on $\eta$ and $\eta'$. For such a representation, presumably a linearization formula can be found and the short-distance singularity extracted as in Minkowski space (since the kernel, even though it depends explicitly on time, has a finite limit as $x' \to x$).

The second important point -- extracting differential operators -- exploits the fact that conservation restricts the form that the stress tensor two-point function can take. For a Poincaré invariant state, taking into account conservation reduces the number of independent functions from five to two, and we extract those operators such that conservation is fulfilled automatically. The remaining two functions are then in principle completely arbitrary, so that one can perform renormalization separately on the most divergent terms which are the only ones that need to be renormalized. It has the further advantage that the strength of the divergence in those functions is reduced to the one of a product of undifferentiated propagators, which need counterterms that diverge like $(n-4)^{-1}$ as appropriate for a one-loop calculation. For the stress tensor correlation function of the massless, minimally coupled scalar, this idea is used in~\cite{fordroman2005,fordwoodard2005} in Minkowski space and in~\cite{parkwoodard2011} in de Sitter space; it also arises quite naturally in Fourier space in a Minkowski background~\cite{martinverdaguer2000}. However, in curved spacetime it is not easy to find the appropriate generalization for the operators $S_{ab}$~\eqref{sab_def}, whose definition was motivated by their symmetry and vanishing divergence. As noted before, $S_{ab}$ with both indices referring to the same point can presumably be substituted by the expansion of the Ricci scalar up to linear order in the metric perturbations $R = S_{ab} h^{ab}$. However, the term $S_{a(c} S_{d)b}$ where the index pairs $(ab)$ and $(cd)$ refer to different spacetime points will generally fail to be divergence-free when acting on a scalar function when generalized in the obvious way by substituting parallel propagators for the metric; it is only in Minkowski space where the parallel propagator coincides with the metric that its divergence vanishes. In~\cite{parkwoodard2011}, the differential operator obtained by the expansion of the Weyl tensor was proposed but yields unwieldy results.\footnote{This includes strong divergences like $(n-4)^{-3}$ in intermediate steps, which however may be an artefact of the concrete calculation --- the final result is the expected one, and the counterterms coincide with well established ones.} From the gauge invariance of the interaction $h_{ab} T^{ab}$ and the fact that the curvature tensors (minus their background value) are gauge-invariant, one may strongly suspect that those differential operators have to come from the expansion of curvature tensors, so that maybe the choice of the Riemann tensor instead of the Weyl tensor will give nicer results. 

The calculation of the stress tensor correlation function for other types of fields can be done with the same ease; since their propagators are obtained by taking derivatives of the scalar propagator, the exactly same Mellin-Barnes representation can be used. The only nontrivial issue (in the author's opinion) is to check if the integrand in the analogue of equation~\eqref{2pf_mellin} is conserved on its own so that the rest of the procedure goes through.

\acknowledgments

It is a pleasure to thank Chris Fewster for stimulating discussions and important references. I would also like to thank the Mathematics Department of the University of York for its hospitality. Financial support through a FPU scholarship no. AP2010-5453 including a short stay at an external research institution, as well as partial financial support by the Research Projects MCI FPA2007-66665-C02-02, FPA2010-20807-C02-02, CPAN CSD2007-00042, within the program Consolider-Ingenio 2010, and AGAUR 2009-SGR-00168, is acknowledged.

\appendix

\section{Curvature tensor expansions}
\label{appendix_expansion}

In this appendix we present the expansion of curvature tensors around a Minkowski background up to linear order in the perturbation.

The perturbed metric and the Christoffel symbols are given by
\begin{equation}
\begin{split}
g_{ab} &= \eta_{ab} + h_{ab} \\
g^{ab} &= \eta^{ab} - h^{ab} \\
\sqrt{-g} &= 1 + \frac{1}{2} h \\
\christoffel{a}{b}{c} &= \frac{1}{2} \left( \partial_b h^a_c + \partial_c h^a_b - \partial^a h_{bc} \right) \eqend{,}
\end{split}
\end{equation}
where the indices on $h_{ab}$ are raised using the background Minkowski metric, and the curvature tensors follow straightforwardly
\begin{equation}
\begin{split}
R^{ab}{}_{cd} &= - 2 \partial^{[a} \partial_{[c} h_{d]}^{b]} \\
R_{ab} &= \frac{1}{2} \left( \partial_a \partial^d h_{bd} + \partial_b \partial^d h_{ad} - \partial_a \partial_b h - \dalembert h_{ab} \right) \\
R &= \partial_a \partial_b h^{ab} - \dalembert h \eqend{.}
\end{split}
\end{equation}

\section{Momentum space regularization and renormalization}
\label{appendix_regren}

In this appendix we show how dimensional regularization and renormalization in position space is related to the usual momentum space techniques. For simplicity, we will treat the massless case, which also is the only term we need to renormalize after extracting differential operators from the stress tensor two-point function. Using the formula~\eqref{feynman_massless} for the Fourier transform of the massless Feynman propagator, we obtain the well-known expression
\begin{equation}
\int \frac{1}{(x^2 + \mathi 0)^{\frac{n}{2}-1}} \mathe^{- \mathi p x} \total^n x = - \mathi \frac{4 \pi^\frac{n}{2}}{\Gamma\left( \frac{n}{2}-1 \right)} \frac{1}{p^2 - \mathi 0} \eqend{.}
\end{equation}
To the product in coordinate space corresponds a convolution in momentum space
\begin{equation}
\begin{split}
\int \frac{1}{(x^2 + \mathi 0)^{\frac{n}{2}-1}} \frac{1}{(x^2 + \mathi 0)^{\frac{n}{2}-1}} \mathe^{- \mathi p x} \total^n x &= - \frac{16 \pi^n}{\Gamma^2\left( \frac{n}{2}-1 \right)} \int \frac{1}{(p-q)^2 - \mathi 0} \frac{1}{q^2 - \mathi 0} \total^n q \\
&= - \mathi \frac{4^{2-\frac{n}{2}} \pi^\frac{n}{2} \Gamma\left( 2-\frac{n}{2} \right)}{\Gamma(n-2)} \frac{1}{(p^2 - \mathi 0)^{2-\frac{n}{2}}} \eqend{,}
\end{split}
\end{equation}
which can be done in $n$ dimensions but is divergent as $n \to 4$. In momentum space one now expands around $n = 4$ to get
\begin{equation}
\int \frac{1}{(x^2 + \mathi 0)^{n-2}} \mathe^{- \mathi p x} \total^n x = \mathi \pi^2 \left( \frac{2}{n-4} - 2 + 3 \gamma - \ln 4 + \ln \pi + \ln (p^2-\mathi 0) \right) + \bigo{n-4} \eqend{.}
\end{equation}
The Fourier transform of the logarithm can be taken using the formula~\eqref{feynman_massless} by rewriting it as
\begin{equation}
\label{fourier_log_feynman}
\begin{split}
\int \ln (p^2-\mathi 0) \mathe^{\mathi p x} \frac{\total^4 x}{(2\pi)^4} &= \dalembert \left[ \lim_{n \to 4} \frac{2}{n-4} \int \left( (p^2-\mathi 0)^{-1} - (p^2-\mathi 0)^{\frac{n}{2}-3} \right) \mathe^{\mathi p x} \frac{\total^n x}{(2\pi)^n} \right] \\
&= \mathi \dalembert \left( \frac{2 \gamma - \ln 4 + \ln (x^2 + \mathi 0)}{4 \pi^2 (x^2 + \mathi 0)} \right) \eqend{,}
\end{split}
\end{equation}
and by using $\dalembert [ 4 \pi^2 (x^2 + \mathi 0) ]^{-1} = \mathi \delta^4(x)$, we get combining all of the above
\begin{equation}
\frac{1}{(x^2 + \mathi 0)^{n-2}} = \mathi \pi^2 \left( \frac{2}{n-4} - 2 + \gamma + \ln \pi \right) \delta^4(x) - \frac{1}{4} \dalembert \left( \frac{\ln (x^2 + \mathi 0)}{x^2 + \mathi 0} \right) + \bigo{n-4} \eqend{.}
\end{equation}
This is of course exactly the result we obtained previously in coordinate space~\eqref{feynman_renormalized}.

\bibliographystyle{JHEP}
\bibliography{literature}

\end{document}